\documentclass[sigconf]{acmart}
\usepackage{listings}
\usepackage[section]{placeins}
\usepackage{array}
\usepackage{vcell}
\usepackage{threeparttable}
\usepackage{multirow}

\usepackage{enumitem}

\usepackage{caption}
\usepackage{subcaption}

\AtBeginDocument{%
  }

\copyrightyear{2025}
\acmYear{2025}
\setcopyright{rightsretained}
\acmConference[SIGCSE TS 2025]{Proceedings of the 56th ACM Technical Symposium on Computer Science Education V. 1}{February 26-March 1, 2025}{Pittsburgh, PA, USA}
\acmBooktitle{Proceedings of the 56th ACM Technical Symposium on Computer Science Education V. 1 (SIGCSE TS 2025), February 26-March 1, 2025, Pittsburgh, PA, USA}
\acmDOI{10.1145/3641554.3701864}
\acmISBN{979-8-4007-0531-1/25/02}

\makeatletter
\gdef\@copyrightpermission{
  \begin{minipage}{0.3\columnwidth}
   \href{https://creativecommons.org/licenses/by-nc-nd/4.0/}{\includegraphics[width=0.90\textwidth]{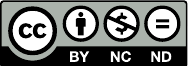}}
  \end{minipage}\hfill
  \begin{minipage}{0.7\columnwidth}
   \href{https://creativecommons.org/licenses/by-nc-nd/4.0/}{This work is licensed under a Creative Commons Attribution-NonCommercial-NoDerivs International 4.0 License.}
  \end{minipage}
  \vspace{5pt}
}
\makeatother

\begin{document}

\title[61A Bot Report: AI homework assistants in CS1]{61A Bot Report: AI Assistants in CS1 Save Students Homework\\ Time and Reduce Demands on Staff. (Now What?)}

\author{J.D. Zamfirescu-Pereira}
\email{zamfi@berkeley.edu}
\orcid{0000-0002-5310-6728}
\affiliation{%
  \institution{UC Berkeley EECS}
  \city{Berkeley}
  \state{California}
  \country{USA}
}

\author{Laryn Qi}
\email{larynqi@berkeley.edu}
\orcid{0009-0001-0424-8746}
\affiliation{%
  \institution{UC Berkeley EECS}
  \city{Berkeley}
  \state{California}
  \country{USA}
}

\author{Bj\"orn Hartmann}
\email{bjoern@eecs.berkeley.edu}
\orcid{0000-0002-0693-0829}
\affiliation{%
  \institution{UC Berkeley EECS}
  \city{Berkeley}
  \state{California}
  \country{USA}
}

\author{John DeNero}
\email{denero@berkeley.edu}
\orcid{0000-0001-9152-3891}
\affiliation{%
  \institution{UC Berkeley EECS}
  \city{Berkeley}
  \state{California}
  \country{USA}
}

\author{Narges Norouzi}
\email{norouzi@berkeley.edu}
\orcid{0000-0001-9861-7540}
\affiliation{%
  \institution{UC Berkeley EECS}
  \city{Berkeley}
  \state{California}
  \country{USA}
}

\begin{abstract}
LLM-based chatbots enable students to get immediate, interactive help on homework assignments, but even a thoughtfully-designed bot may not serve all pedagogical goals. We report here on the development and deployment of a GPT-4-based interactive homework assistant (``61A Bot'') for students in a large CS1 course; over 2000 students made over 100,000 requests of our Bot across two semesters. Our assistant offers one-shot, contextual feedback within the command-line ``autograder'' students use to test their code. Our Bot wraps student code in a custom prompt that supports our pedagogical goals and avoids providing solutions directly. Analyzing student feedback, questions, and autograder data, we find reductions in homework-related question rates in our course forum, as well as reductions in homework completion time when our Bot is available. For students in the $50^{th}-80^{th}$ percentile, reductions can exceed 30 minutes per assignment, up to 50\% less time than students at the same percentile rank in prior semesters. Finally, we discuss these observations, potential impacts on student learning, and other potential costs and benefits of AI assistance in CS1.
\end{abstract}



\maketitle

\section{Introduction}

The recent wide availability of Large Language Models (LLMs) has given students in introductory Computer Science (CS) courses a tempting alternative to asking a human TA for help on programming assignments---and potentially waiting hours to receive it. For instructors of large courses, LLMs like ChatGPT appeal by offering 24/7 access to personalized assistance that will answer nearly any student question, though not always correctly, and with other risks. 

But while naively used LLMs do help students solve assigned problems, they typically do so by providing correct answers along with explanations, allowing students to avoid the process of developing solutions themselves and the learning associated with this process. A number of recent reports~\cite{denny2023chat, denny2023computing, finnie2023my, hellas2023exploring, kazemi2024codeaid, liffiton2023codehelp, wang2024large} present more thoughtful approaches: new systems, also based on LLMs, geared towards offering guidance and assistance without providing direct solutions. 

Both students and instructors are reported to find these systems helpful~\cite{kazemi2024codeaid, liffiton2023codehelp}. Recent studies of deployed LLM-based assistants have included analyses of what affordances are most appreciated and what kinds of functionality students are most likely to use~\cite{kazemi2024codeaid,denny2024desirable}; how to effectively design systems with guardrails to reduce instances of solution-sharing~\cite{liffiton2023codehelp}; and what kinds of error messages are most correlated with reductions in error rates~\cite{wang2024large}. There is a growing interest in understanding the landscape of these new systems and their impacts on students learning computer science~\cite{prather2024instructors}---but while much is known about how these systems are designed and what seems to work well, comparatively less has been reported on how these systems impact specific courses.

In this experience report, we first present our own assistant, a \textit{low-friction} ``61A Bot'' with 24/7 availability that offers feedback \textit{on every run} of an ``autograder'' that students use liberally to test their code-in-progress. Our Bot constructs a GPT-4 request using our custom prompt, homework question text, student code, and autograder error output (where available), returning its response to students. The prompt is itself designed to steer towards feedback that mirrors how we ourselves typically approach student questions, aligned with recent work in this area~\cite{markel2021inside}: identifying whether the student understands the question, which concepts students might need reinforcement on, and whether they have a plan, and then helping students by providing conceptual, debugging, or planning support as appropriate.

Then, we examine our Bot's impact on our CS1 course, UC Berkeley's CS 61A---a large ($\sim{}1000$-student) Python-based course targeted at CS majors. Ultimately, we ask: \textbf{What impact does this deployment have on our course, on students and on staff?} First, we find that students assess a majority of Bot-provided hints as being helpful, in line with prior work. Second, through a retrospective observational study aimed at quantifying impacts, we find that requests for homework help to our course forum drop dramatically (75\%) after our Bot is deployed, and that students spend substantially less time completing their homework---25-50\% less time, often more than 30 minutes faster, for students at the same percentile rank in completion time across prior semesters.

We note that while these impacts do not necessarily imply improvements in student learning outcomes, speeding up homework completion times without increasing learning could still be a positive outcome: the time saved on completing traditional assignments could be harnessed by instructors to increase learning in other ways. In our final section, we discuss some of the costs and potential benefits of deploying LLM-based assistance in CS1, exploring possible learning impacts, reductions in the visibility of student challenges to staff, and opportunities for future research.

Our work makes three main contributions. First, it offers a rich description of our low-friction CS1 assistant and its deployment in our course. Second, we detail findings of students' subjective experiences based on student surveys over two semesters. Third, we share results from a retrospective, observational analysis of student homework completion patterns, identifying quantitative impacts our Bot had on staff demand and students' homework experiences.

\section{Background \& Related Work}

Generative models such as ChatGPT,\footnote{\url{https://chat.openai.com/}} OpenAI Codex~\cite{chen2021evaluating}, 
Amazon CodeWhisperer,\footnote{\url{https://aws.amazon.com/codewhisperer/}} and GitHub Copilot\footnote{\url{https://copilot.github.com/}} offer promising opportunities for enriching the learning experience of students. These models have already been leveraged by educators in different areas of Computing education~\cite{finnie2022robots, denny2023computing, hellas2023exploring, denny2023chat}, where they accelerate content generation and seem to be impacting the relevant skills students gain in introductory CS courses. Researchers have studied LLMs in areas such as generating code explanations~\cite{leinonen2023comparing, becker2023programming, macneil2023experiences, harvardcs50}, providing personalized immediate feedback~\cite{bassner2024iris}, enhancing programming error messages~\cite{leinonen2023using, wang2024large}, generating discussion forum responses~\cite{mitra2024elevating, liu2024teaching}, and automatic creation of personalized programming exercises and tutorials~\cite{sarsa2022automatic, yuan2023evaluating, reeves2023evaluating} to enhance the comprehensiveness of course materials.

However, the integration of LLMs in CS1 instruction comes with challenges. Students could become overly reliant on automation (a concern at least as old as calculators~\cite{demana2000calculators}), potentially hindering their development of critical problem-solving skills---though recent work suggests these negative effects can be avoided, at least for programming assistance~\cite{kazemitabaar2023studying}. Taken to an extreme, the resulting absence of human interaction could have negative effects, alongside other ethical concerns related to plagiarism and the responsible use of LLM-generated code. To maximize the benefits of LLMs while mitigating these challenges, a thoughtful and balanced approach to their incorporation into CS1 courses is essential~\cite{mirhosseini2023your, macneil2023implications, denny2024desirable}.

Through deployments of LLMs as intelligent tutors, students can receive immediate, personalized support and guidance, which would ideally foster a deeper understanding of coding concepts and promote self-paced learning---just as with pre-LLM Intelligent Tutoring Systems (e.g.,~\cite{suzuki2017tracediff}; see~\cite{crow2018intelligent} for a review). The ability of LLMs to generate \textit{tailored} resources, such as new, personalized tutorials and newly-generated code examples, not only expands the available learning materials but also accommodates students' varying learning preferences---though these generated materials are not always better~\cite{pardos2023learning}. Educators should integrate LLMs as \textit{complementary} tools, striking a balance between automation and human interaction while emphasizing the development of critical problem-solving skills and responsible coding practices, ultimately serving students better in their CS education. 

Researchers are also increasingly integrating LLM-based \textit{chatbots} in courses~\cite{harvardcs50, wu2022student} and online educational websites~\cite{ofgang2023khanmigo} to provide immediate personalized feedback, and in tools in supporting students' development of programming skills~\cite{prather2023s, finnie2023my, cipriano2023gpt}. These include CodeHelp~\cite{liffiton2023codehelp, sheese2024patterns} and CodeAid~\cite{kazemi2024codeaid}, two systems (and deployments) that bear a number of similarities to our own---though those systems enable students to ask questions, while ours (we believe uniquely) integrates feedback directly into the tool students already use to execute that code, and then builds on the student's history of prior assistant hints and code changes in response.

Our work here builds on these prior efforts in the design of 61A Bot, differing primarily in its integration mode (embedded within an autograder students must run anyway), and the lack of conversational interaction. Our evaluation validates prior findings on engagement and satisfaction, and additionally offers a unique perspective of the impact of our Bot on student assignment completion times compared with historical course baselines.

\section{Design \& Deployment}

Our deployment focused on providing students help with homework problems in part to address frequent student feedback from prior terms about long wait times for TA support for homework problems. In particular, we chose to focus primarily on the kinds of debugging assistance our staff are often asked for. 
Three concerns---hallucinations, students sharing personal information with a third party, and the harms from an unmonitored chat interaction---led us to a one-shot ``Get Help'' design rather thanc``chat''.
This meant eschewing the valuable pedagogical approach of having students explain their understanding of the problem.

Following a common tutoring pattern~\cite{markel2021inside}, we designed a prompt that would try to assess student conceptual knowledge, based on the provided code, and offer syntactical, logical, or even template-code suggestions---but not solutions. This prompt explicitly includes a sequence of questions to consider in response to the student code: 
\begin{enumerate}
    \item Is the student missing conceptual knowledge?
    \item Is their current code on the right track?
    \item How close are they to a solution?
    \item Were they able to follow previous advice? \label{q4}
    \item Do they have a reasonable plan?
\end{enumerate}
Though we avoided students explicitly writing natural language ``chat'' messages to the bot, we did want some degree of continuity---which we achieved by also including up to three prior \textit{(student code, Bot advice)} exchanges, if available (enabling question~\ref{q4}).

In addition to the steps above, the prompt also includes a per-problem instruction block, which we used for about 10\% of problems, and more general instructions such as \texttt{Do not give the student the answer or any code.} and \texttt{Limit your response to a sentence or two at most.} 
Our full prompt, along with our server and VS Code extension, are open source and available online at \underline{\url{https://github.com/zamfi/61a-bot}}.

\begin{figure}[!t]
    \centering
            \includegraphics[width=0.98\linewidth]{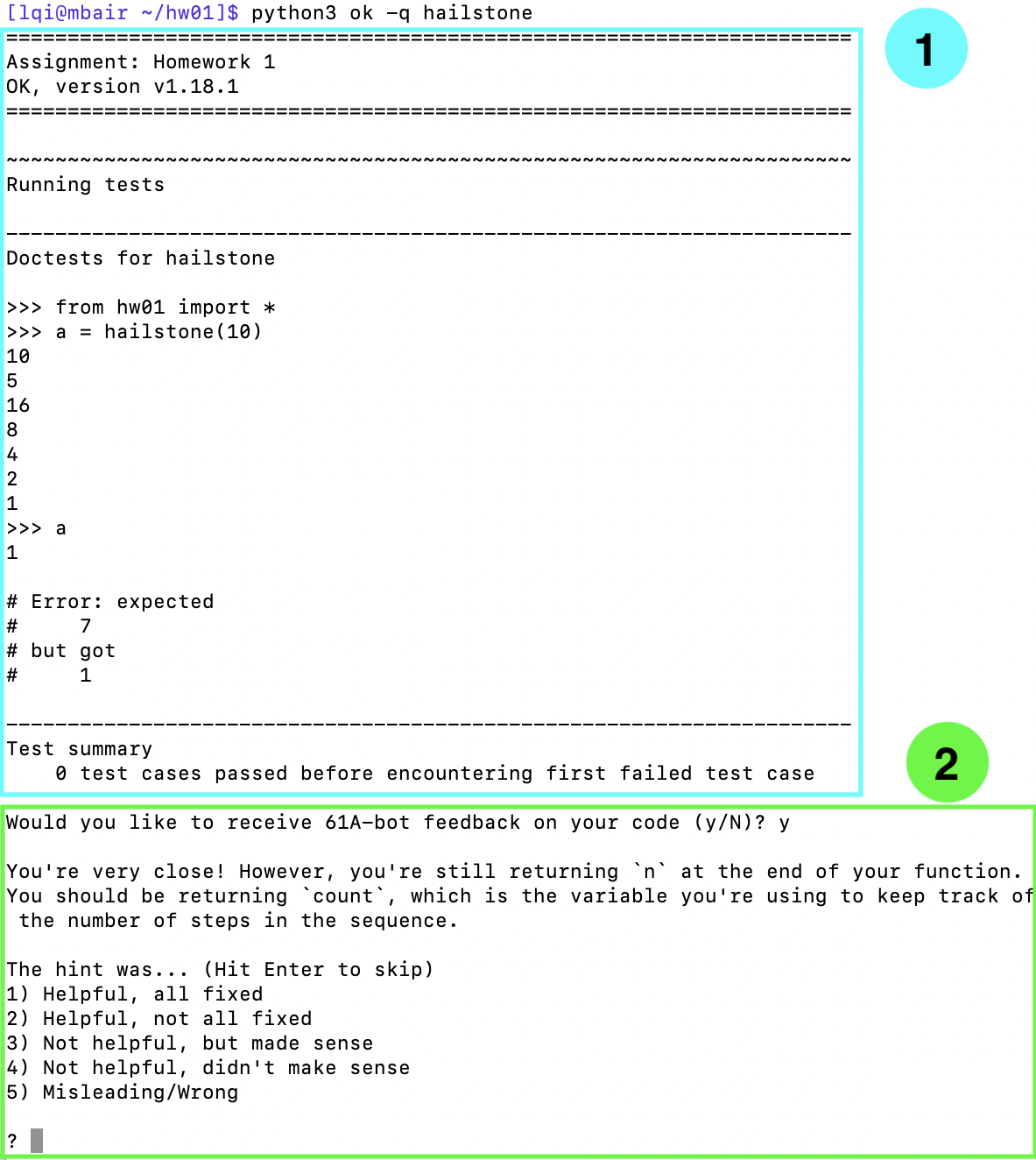}
        \caption{Screenshot of the ``autograder'' command-line interface. Students run the autograder to test their code against a test suite \textbf{(1)}; if any test cases fail, they are asked if they would like to receive Bot feedback, and then asked to assess the hint \textbf{(2)}.     }
    \label{fig:extension-screenshot}
\end{figure}

\subsection{Course Details \& Deployment} \label{sec:deployment}

Our course covers most of the typical CS1 content, plus a few additional topics, and we report here on the CS1 portion of our course. These modules are taught using Python, to a student population made up predominantly of CS majors and intended majors ($50\%\pm10\%$~majors/intended), with substantial prior preparation ($80\%\pm{}5\%$~any prior CS course, including high school). These demographics are broadly consistent across the semesters we examine historically in this report, and variations do not correlate with the results described here. (Typical Fall and typical Spring semester students differ demographically due to how course requirements are structured at our institution; we treat those as separate populations.) In our course, content is typically introduced in lecture, and then reinforced in three assignment types: first, ``labs,'' which include a mixture of notional machine reinforcement (e.g., tracing, ``What would Python do?'') and traditional homework problems, completed in small groups without access to 61A Bot. These are followed by ``homework,'' which students complete individually \textit{with} access to 61A Bot. Finally, students complete ``projects'' that bring together multiple concepts towards a single goal, over a few weeks, again without access to 61A Bot. In addition, students were prohibited by course policy from using ChatGPT or other similar AI-based systems for help across all assignment types---but we do not have the means to enforce this.

\subsubsection{Using 61A Bot}

Students primarily (93\%) access our Bot through an autograder\footnote{
OK Client, \url{https://github.com/okpy/ok-client}
} in the command line which provides the result of running the student's code on a set of test cases; if any fail, the student is then asked whether they would like to receive feedback on their code from 61A Bot (see Figure~\ref{fig:extension-screenshot}). If they do, the autograder collects the student's code, any errors from executing test cases, and constructs a request from these.
Alternatively, students choosing to use VS Code\footnote{Visual Studio Code, \url{https://code.visualstudio.com}} can install an extension to enable a ``Get Help'' button in the toolbar that works similarly.

Requests thus include: our prompt, along with any problem-specific notes; the text of the specific homework problem the student is requesting help for (one of 4-6 problems within an assignment); up to 3 pairs of prior-code/Bot-response text (from any prior requests); the student's current code; and, finally, any error text from failed test cases.
These requests are sent to a server run by our instructional staff, which passes it to GPT-4 and logs the request and GPT-4's response for further analysis. 

Students are informed when installing the software that, in using our assistant, all code they write will be sent to OpenAI via Microsoft Azure,\footnote{\url{https://ai.azure.com/}} and that they should not include any content in their code files (e.g., comments) that they do not want to share.

\subsubsection{Deployment Timeline} \label{sec:deployment-timeline}

We piloted and continued development on 61A Bot throughout the academic year 2023-2024. In Fall 2023, we deployed an initial pilot of 61A Bot to a section of 400 students. This was followed by a full-scale deployment in week 10 (after the CS1 portion of our course) for the approximately 1400 students across both sections of the course. 
In conjunction with the wider deployment, we also enabled access through the autograder tool students could already run from the command line to validate their code against a set of test cases. In Spring 2024, all 900 enrolled students had access to both modalities of the Bot from the start of the semester. Our qualitative analysis thus draws on both semesters, while our quantitative analysis is focused on the full deployment in Spring 2024 where we can compare like-for-like with a full-term deployment and historically comparable student populations.

\section{Outcomes} \label{sec:findings}

Students' adoption of 61A Bot was immediate, and usage exploded once we integrated access into our autograder. In this section, we detail usage patterns and changes in reliance on course staff, report on student assessments of their experiences, and finally offer an investigation of the possible impacts of 61A Bot on homework completion times. Unless otherwise noted, statistical accounts in this section come exclusively from Spring 2024, the semester in which our fully-developed Bot was  deployed for all students (see \S\ref{sec:deployment-timeline} for deployment details).

\subsection{Usage \& Reliance on Staff} \label{sec:usage}
Usage patterns show that students are returning to 61A Bot multiple times as they engage in homework: across our pilot and full deployment semesters, over 2000 students made a total of 105,689 requests of our bot. The median student in our full deployment semester made 25 requests to our bot, rising to 80 requests for the student at the 95th percentile. As expected, usage increases as the assignment deadline nears and is concentrated in the late afternoons and evenings---a pattern similar to the usage reported in~\cite{liffiton2023codehelp,kazemi2024codeaid}---with a peak request rate of 291 requests/hour.

This engagement correlates with a reduction in help requests on 
our online discussion forum for students to receive asynchronous help. There is a substantial (30\%) decrease in the number of questions asked (scaled to total enrollment) throughout the semester from Spring 2023 to Spring 2024, from 1741 to 1185 questions per thousand students. The impact on homework-specific questions is even larger, showing a 75\% decrease from 344 homework questions per thousand students in Spring '23 to 88 in '24---see \autoref{tab:ed-questions}. (We include scaled question counts for Fall 2022 and Fall 2023 as data points illustrating the level of consistency across semesters.)

\begin{table}[!t]
    \centering
            \begin{tabular}{p{0.13\textwidth}p{0.05\textwidth}p{0.05\textwidth}p{0.05\textwidth}|p{0.05\textwidth}}
        \toprule
        {} & \textbf{Fa22} & \textbf{Sp23} & \textbf{Fa23} & \textbf{Sp24} \\
        \midrule
        \textbf{\# Students} & 1656 & 1169 & 1407 & 872 \\
        \midrule
        \textbf{\# Questions} & 1853 & 2035 & 1823 & 1033 \\
        \textit{\hspace{0.7em}per 1000 students} & 1119 & 1741 & 1296 & 1185 \\
        \midrule
        \textbf{\# HW Questions} & 234 & 402 & 177 & 77 \\
        \textit{\hspace{0.7em}per 1000 students} & 141 & 344 & 126 & 88 \\
        \bottomrule
    \end{tabular}
    \caption{Average number of forum questions, Fall 2022--Spring 2024, for the CS1 portion of the class (Fall 2023 results thus do not include the period of widespread deployment).}
    \label{tab:ed-questions}
\end{table}

\subsection{Student Reception} \label{sec:student-feedback}

Student feedback suggests that students also found the Bot helpful. We solicited this feedback in two ways: First, we queried students for their assessment of each individual hint, which we received for approximately 27\% ($7459 / 27419$) of queries.\footnote{These counts reflect queries from only those students who gave consent for their data to be used for research, and only during the CS1 portion of our course.} Of these, 70\% ($5210 / 7459$) were rated as ``helpful,'' with 45\% of those, $2368/5210$, reporting that the problem was now resolved. A further 10\% ($743 / 7459$) were rated as ``not helpful, but made sense,'' while the final 20\% ($1496 / 7459$) were rated as insensible, misleading, or wrong.

Second, we formally surveyed students on their usage and perceptions of the Bot. In Fall 2023, we conducted a non-anonymous survey at the end of the semester to which 49\% ($698/1407$) of students responded. Students were asked to rate how much they used the Bot and how helpful they found it on a scale from 1 to 5.  As expected, those who reported more usage also found it more helpful. 

\begin{figure}
    \centering
    \includegraphics[width=0.99\linewidth]{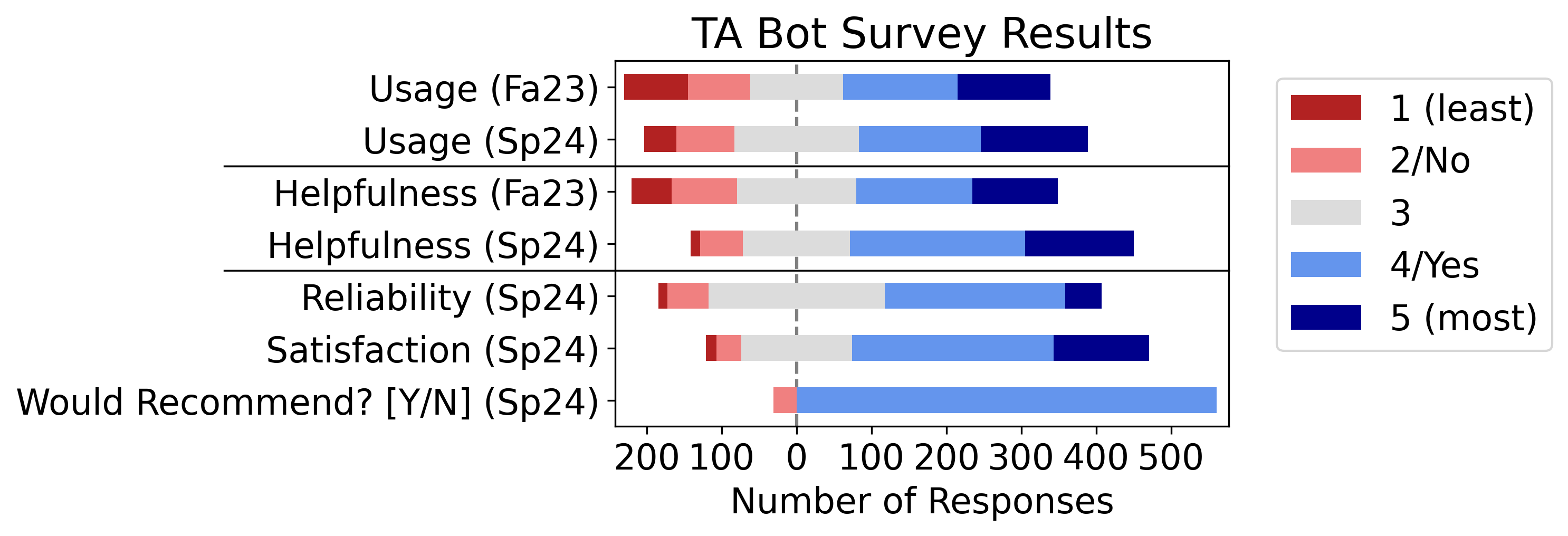}
    \caption{61A Bot Survey Results. Students were asked to report on their usage and perceptions of Bot helpfulness and reliability, as well as their overall satisfaction on a scale of 1-5, from least to most.}\label{tab:survey-results}
\end{figure}

In Spring 2024, we conducted a non-anonymous survey at the end of the semester, to which 89\% ($774/872$) of students responded. On a scale from 1 to 5, we asked students to rate their Bot usage, Bot helpfulness, Bot reliability, and overall satisfaction with the Bot. Finally, we asked them whether or not they recommend that the Bot be available to students in future semesters. The results from these surveys can be found in \autoref{tab:survey-results}.

Note that these results include both our partial- (Fall 2023) and full-deployment (Spring 2024) semesters; the Fall 2023 results thus reflect usage only post-deployment. Additionally, these surveys were non-anonymous to track individual participation (for pedagogical goals independent of this project); because of this non-anonymity, however, we did not ask whether students relied on ChatGPT or other prohibited (by course policy) methods of support, reasoning that we were unlikely to be able to rely on such results.

In the Spring 2024 survey, we also asked two optional free-response questions: What did students like the most and least about learning with the Bot? We include a representative response for each here: 
\underline{Most liked}:

\noindent
    ``What I loved about the bot is that it allowed me to get feedback when I didn't have access to a tutor. Accordingly, instead of banging my head against the wall for hours, I was able to get feedback about what I was doing wrong and correct the mistakes. For me personally, I would have had a lot more success in this class if I would have had access to the bot for labs. (Labs on average took me about 2 and a half hours to complete, and sometimes longer if I didn't have access to a tutor).''

\underline{Least liked}:

\noindent
    ``Sometimes the answers were slightly vague. Of course, the bot can't simply spit out the answer, but sometimes it was frustrating how it would say `you're on the right track, but there seems to be a conceptual misunderstanding with \_\_' --- the explanations for the blank could be a bit jargon-filled and didn't always directly help me resolve the misunderstanding due to imprecise language.''

Students generally appreciated 61A Bot's accessibility, debugging capabilities, and time savings. However, the hints were sometimes too vague for the students to make changes, while other times, the Bot was too specific and gave away too much. 
Despite our attempts to steer GPT-4 towards rephrasing and a broader diversity of responses to repeat inputs, these incidents still occur. 
\subsection{Effects on Homework} \label{sec:completion-time}

To understand the effects of 61A Bot on students in our course, we compare student performance from our semester of full deployment, Spring 2024 (SP24), with performance from prior Spring semesters (see~\S\ref{sec:deployment}), going back through Spring 2021 (SP21)---comprising a total of 1,643,613 data points from 6,034 students. (Our IRB does not allow differential access to tools in courses, preventing a randomized control trial.)

Though 61A Bot is available to students for \textbf{homework} assignments, it is not available for \textbf{lab} assignments nor for \textbf{projects} (see \S\ref{sec:deployment}). To the extent that performance differences on homework assignments between pre- and post-deployment semesters are inconsistent with performance differences on lab assignments and projects, some of this difference may be attributable to the use of the Bot. There is some variation in course staff, individual lectures, and specific problems within labs and homework assignments over time in our course; where problems have changed more than trivially in content or sequencing, we have omitted them from our comparisons. The specific assignments we report on here are otherwise representative of the full set of assignments.

\begin{figure}[!b]
    \centering
    \begin{subfigure}[b]{\linewidth}
        \centering
        \includegraphics[width=0.49\linewidth]{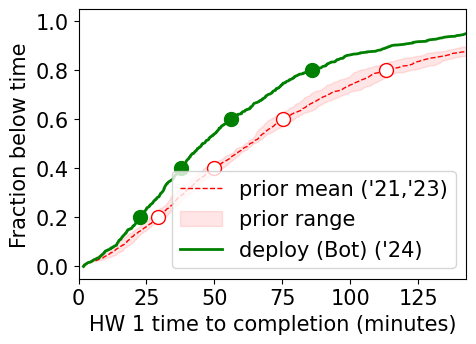}
        \includegraphics[width=0.49\linewidth]{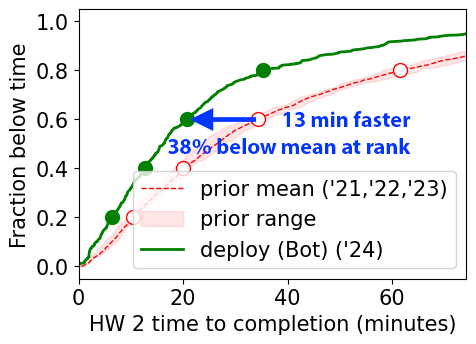}
        \includegraphics[width=0.49\linewidth]{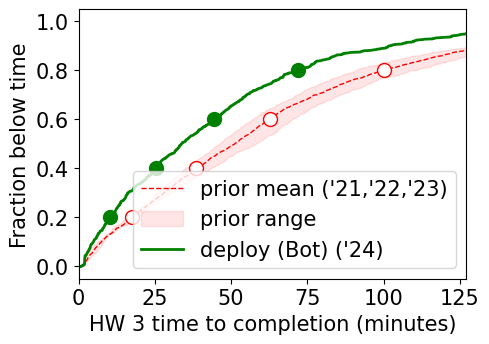}
        \includegraphics[width=0.49\linewidth]{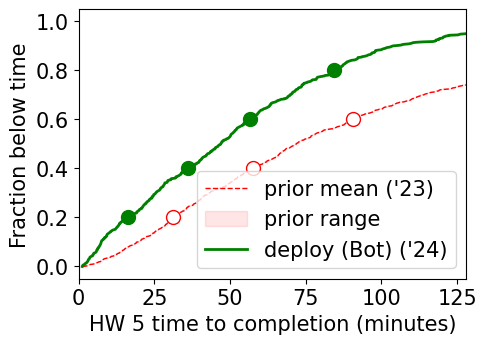}
        \caption{HW assignments $1^*$, $2^*$, $3^*$, and $5^*$.\vspace{1em}}
        \label{fig:hw-cdfs}
    \end{subfigure}
    
    \begin{subfigure}[b]{\linewidth}
        \centering
        \includegraphics[width=0.49\linewidth]{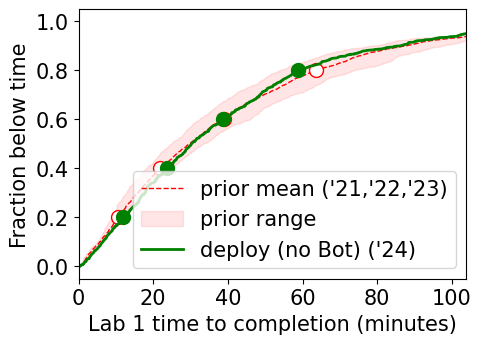}
        \includegraphics[width=0.49\linewidth]{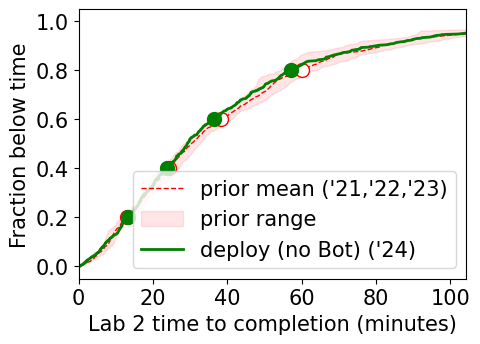}
        \includegraphics[width=0.49\linewidth]{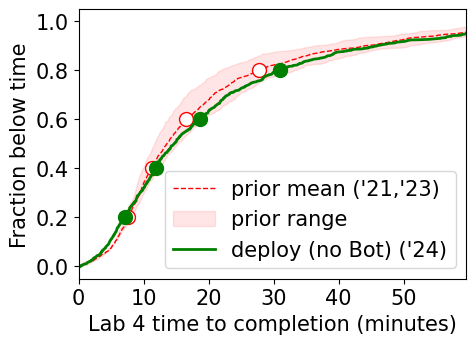}
        \includegraphics[width=0.49\linewidth]{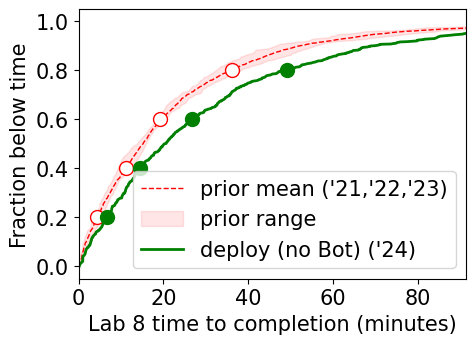}
        \caption{Lab assignments $1$, $2$, $4$, and $8^*$.\vspace{1em}}
        \label{fig:lab-cdfs}
    \end{subfigure}

    \begin{subfigure}[b]{\linewidth}
        \centering
        \includegraphics[width=0.49\linewidth]{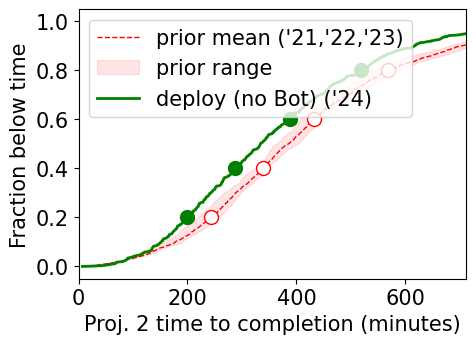}
        \includegraphics[width=0.49\linewidth]{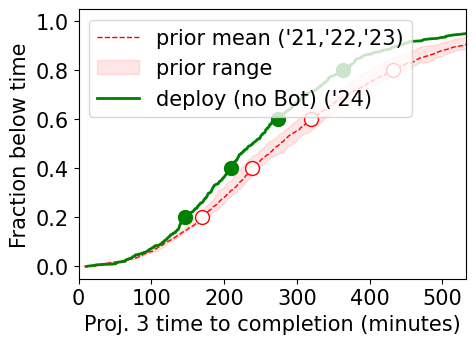}
        \caption{Projects $2^*$ and $3^*$.}
        \label{fig:project-cdfs}
    \end{subfigure}
    \caption{CDFs of assignment completion times. Circles identify the 20, 40, 60, and 80th percentiles; differences can be read by matching circle markers horizontally. E.g., for HW2, the mean 60th percentile time across SP21-SP23 is 34 minutes; in SP24 this time is 21 minutes, a 38\% reduction.\\
    $^*$ denotes differences significant at the $p < 0.001$ level.}
    \label{fig:cdfs}
\end{figure}

\subsubsection{Data \& Analytical Method} \label{sec:data-analysis}

Our course ``autograder'' records every attempt a student makes to test their code, storing a ``snapshot'' of student code on our instructional servers, along with a student identifier and a timestamp; we use these to reconstruct a history of student progress. Autograder use is unlimited, and students typically revise their programs repeatedly until they are ready to submit their final code. Nearly all students submit code that successfully passes all test cases, so examining the final submitted code artifact alone does not necessarily provide useful insight into impacts. Thus the metric we consider here is \textit{time to completion}, rather than passing test cases or other measures of code quality. 

We calculate an approximate total completion time by summing the timestamp deltas between snapshots. To account for students completing the homework across multiple sessions, we ignore time deltas above a 60-minute threshold---a value chosen somewhat arbitrarily, but we confirm that the results we report here are robust to values in the range of 20 to 120 minutes.\footnote{An independent analysis of snapshot \textit{counts} were consistent with the findings we report here, as expected from~\cite{piech2012modeling}; we report times as we found them more straightforward to reason about.}
For clarity, we aggregate individual repeated problems into ``assignments,'' e.g., ``HW 1.'' Occasionally, one assignment in a particular semester differs sufficiently from other semesters that we omit it entirely.\footnote{We chose to aggregate in this way for clarity; our reported analyses are robust to aggregation method and problem selection, as well as many other factors elided here for space.}

We then report the distribution of student \textit{assignment completion times} using Cumulative Distribution Function (CDF) plots for specific homework assignments, labs, and projects. This CDF can be read as ``What fraction of students (y-axis) complete the homework in less than some number of minutes (x-axis)?'', capturing how long students take to complete these assignments across the full distribution of completion times.

\subsubsection{Results}

Our primary finding is that student completion times on identical \textit{homework} assignments are substantially faster in our post-deployment semester, Spring 2024, compared to prior semesters (\autoref{fig:hw-cdfs}). This effect is not seen in the similar \textit{lab} assignments where the Bot is not available (\autoref{fig:lab-cdfs}); a smaller, but still substantial, speedup is seen in \textit{projects} (\autoref{fig:project-cdfs}).

The data indicate that the (Bot-available) homework completion times in Spring 2024 are between $25\%-50\%$ faster among students in the middle $20-80^{th}$ percentile of completion times, when compared to students at the same percentile rank (\autoref{fig:hw-cdfs}). There is some inherent variability in completion time from year to year for a given percentile rank (e.g., the $80^{th}$ percentile ranked student); the red bands in \autoref{fig:cdfs} show this variability. Note that the Spring 2024 homework completion times within the $20-80^{th}$ percentile range were lower, at a given percentile rank, than the mean completion time at that rank across prior semesters---by a factor of $3-9\times$ the range of that prior variability across Spring semesters.
In contrast, no-Bot \textit{lab} completion times fall within the range of the prior semesters' variability for the first few labs. Most later labs lack enough consistency across semesters for a robust comparison, but lab 8 does offer a glimpse into one possible outcome: here, students in Spring 2024 at a given rank took \textit{more} time than students at the same rank in any prior semesters.

Meanwhile, the no-Bot \textit{project} completion times fall somewhere in between: Spring 2024 students completed these projects between $10\%-20\%$ faster than the pre-deployment mean---a speedup about half as large as the Bot-available homeworks.

\section{Discussion \& Future Work}

Overall, our findings are consistent with a causal link between the availability of our Bot and faster homework completion times. In the context of our course (see \S\ref{sec:data-analysis}), reduced completion times are likely to be a result of students reaching a correct solution more quickly, rather than stopping before reaching a correct solution, or making other mistakes, as might be the case in domains where assignments have a greater variety of possible outcomes.

Though we did not carefully examine (and don't make claims about) student learning outcomes, we have reason to believe that students are not performing much worse after deployment. In particular, student exam performance is sufficiently inconsistent (exam coverage differs from term to term) that we did not report on it here, but it does not appear that Spring 2024 students performed much worse than in prior terms. Additionally, the differences in performance on assignments where the Bot was not available (labs and projects) were much smaller. We can't say for certain, but we have not found much evidence for a major decrease in outcomes.

If indeed this reduction in homework time results in little or no learning loss, we can then ask: What implications does this have for CS1 courses? Are there other costs unrelated to learning loss? And what other benefits might accrue?

On the costs side, with most homework help requests going through an LLM, our human TAs may struggle more to stay on top of common challenges among students, leading to a looser feedback loop. However, this challenge could be addressed with new tools that help TAs aggregate over automated help requests---but now with LLM-generated hints and student feedback on whether those hints are helpful. That is, of course, assuming CS1 courses do not elect to reduce staff in response.
Given the mode of access of 61A Bot, one likely change in student learning is a reduction in the ability to understand error traces. With an easy way to get natural language feedback on failed test cases, students are less incentivized to read traces, let alone understand them.

However, instructors could use this extra student time to cover more material, such as debugging techniques or reading traces. Or, students could continue to simply spend less time on the course. Similarly, TAs could spend less time debugging straightforward homework errors and more time focused on other forms of support.

These factors all point to a need for more research to better understand the actual costs and to inform decisions made in the hope of realizing actual benefits beyond saved student and TA time.

We ourselves plan to also explore new opportunities to extend 61A Bot in future work, in particular in tracking student progress over time, and in providing concept support though worked examples and links to our online textbook.

\subsection{Limitations \& Threats to Validity}

Differences in prior preparation could explain why students complete homework more quickly in Spring 2024---but any such differences would also have to explain the \textit{lack} of a decrease in lab completion times. In fact, the consistency in early lab completion times over the 4 semesters we examined suggests that our student populations do not differ significantly in prior ability.

Similarly, differences in course content delivery, staffing, structure, etc., would be expected to impact both labs and homeworks, as lectures and discussion sections for given topics come before labs and homeworks. No additional homework support was provided in Spring 2024 beyond 61A Bot---no hints or support unique to Spring 2024 were offered in lecture or group sessions.

Finally, our study is entirely observational, and there could be other causes for the effects we observe---perhaps students are using ChatGPT for their homework or projects despite the prohibition on ChatGPT use and the availability of 61A Bot. (ChatGPT was originally released in November 2022, but we observed no similar effect in Spring 2023 compared with prior terms.) 

\section{Conclusions}

Our results suggest that 61A Bot reduces demands on staff and helps students complete homework more quickly, with oversized impacts for students who spent the most time on homework---a benefit that might even disproportionately support goals towards inclusion in CS. But, ideally, this type of scaffolding should recede over time as learners become more confident~\cite{soloway1994learner}. 61A Bot has not yet clearly achieved this goal.

Guidelines around the inclusion of AI-based course materials and tools suggest that these should only be incorporated when we have a good understanding that their benefits outweigh their costs~\cite{bala2023generative}. Yet even if the primary outcome of 61A Bot is limited to a reduction in homework completion time with no other benefits, we believe that 61A Bot clears this bar---but that further research into improving outcomes and mitigating the costs we have started to expose is critical and \textit{urgent.}

\begin{acks}

This work was made possible by a few generous sources of support: an Inclusion Research Award from Google, and support for 61A Bot's use of Azure's OpenAI API by Microsoft. We would also like to extend our graitiude to our anonymous reviewers for their helpful feedback, as well as to our student participants in this experiment.

\end{acks}

\clearpage
\bibliographystyle{ACM-Reference-Format}
\bibliography{ref,csbot}


\begin{thebibliography}{38}


\ifx \showCODEN    \undefined \def \showCODEN     #1{\unskip}     \fi
\ifx \showDOI      \undefined \def \showDOI       #1{#1}\fi
\ifx \showISBNx    \undefined \def \showISBNx     #1{\unskip}     \fi
\ifx \showISBNxiii \undefined \def \showISBNxiii  #1{\unskip}     \fi
\ifx \showISSN     \undefined \def \showISSN      #1{\unskip}     \fi
\ifx \showLCCN     \undefined \def \showLCCN      #1{\unskip}     \fi
\ifx \shownote     \undefined \def \shownote      #1{#1}          \fi
\ifx \showarticletitle \undefined \def \showarticletitle #1{#1}   \fi
\ifx \showURL      \undefined \def \showURL       {\relax}        \fi
\providecommand\bibfield[2]{#2}
\providecommand\bibinfo[2]{#2}
\providecommand\natexlab[1]{#1}
\providecommand\showeprint[2][]{arXiv:#2}

\bibitem[Bala et~al\mbox{.}(2023)]%
        {bala2023generative}
\bibfield{author}{\bibinfo{person}{Kavita Bala}, \bibinfo{person}{Alex Colvin}, \bibinfo{person}{Morten~H. Christiansen}, \bibinfo{person}{Allison~Weiner Heinemann}, \bibinfo{person}{Sarah Kreps}, \bibinfo{person}{Lionel Levine}, \bibinfo{person}{Christina Liang}, \bibinfo{person}{David Mimno}, \bibinfo{person}{Sasha Rush}, \bibinfo{person}{Deirdre Snyder}, \bibinfo{person}{Wendy~E. Tarlow}, \bibinfo{person}{Felix Thoemmes}, \bibinfo{person}{Rob Vanderlan}, \bibinfo{person}{Andrea~Stevenson Won}, \bibinfo{person}{Alan Zehnder}, {and} \bibinfo{person}{Malte Ziewitz}.} \bibinfo{year}{2023}\natexlab{}.
\newblock \bibinfo{title}{{Generative} {Artificial} {Intelligence} for {Education} and {Pedagogy} {\textbar} {Center} for {Teaching} {Innovation}}.
\newblock
\newblock
\urldef\tempurl%
\url{https://teaching.cornell.edu/generative-artificial-intelligence/cu-committee-report-generative-artificial-intelligence-education}
\showURL{%
\tempurl}


\bibitem[Bassner et~al\mbox{.}(2024)]%
        {bassner2024iris}
\bibfield{author}{\bibinfo{person}{Patrick Bassner}, \bibinfo{person}{Eduard Frankford}, {and} \bibinfo{person}{Stephan Krusche}.} \bibinfo{year}{2024}\natexlab{}.
\newblock \showarticletitle{Iris: An AI-Driven Virtual Tutor for Computer Science Education}. In \bibinfo{booktitle}{\emph{Proceedings of the 2024 on Innovation and Technology in Computer Science Education V. 1}} (Milan, Italy) \emph{(\bibinfo{series}{ITiCSE 2024})}. \bibinfo{publisher}{Association for Computing Machinery}, \bibinfo{address}{New York, NY, USA}, \bibinfo{pages}{394–400}.
\newblock
\showISBNx{9798400706004}
\urldef\tempurl%
\url{https://doi.org/10.1145/3649217.3653543}
\showDOI{\tempurl}


\bibitem[Becker et~al\mbox{.}(2023)]%
        {becker2023programming}
\bibfield{author}{\bibinfo{person}{Brett~A Becker}, \bibinfo{person}{Paul Denny}, \bibinfo{person}{James Finnie-Ansley}, \bibinfo{person}{Andrew Luxton-Reilly}, \bibinfo{person}{James Prather}, {and} \bibinfo{person}{Eddie~Antonio Santos}.} \bibinfo{year}{2023}\natexlab{}.
\newblock \showarticletitle{Programming Is Hard--Or at Least It Used to Be: Educational Opportunities And Challenges of AI Code Generation}.
\newblock  (\bibinfo{year}{2023}).
\newblock


\bibitem[Chen et~al\mbox{.}(2021)]%
        {chen2021evaluating}
\bibfield{author}{\bibinfo{person}{Mark Chen}, \bibinfo{person}{Jerry Tworek}, \bibinfo{person}{Heewoo Jun}, \bibinfo{person}{Qiming Yuan}, \bibinfo{person}{Henrique Ponde de~Oliveira Pinto}, \bibinfo{person}{Jared Kaplan}, \bibinfo{person}{Harri Edwards}, \bibinfo{person}{Yuri Burda}, \bibinfo{person}{Nicholas Joseph}, \bibinfo{person}{Greg Brockman}, {et~al\mbox{.}}} \bibinfo{year}{2021}\natexlab{}.
\newblock \showarticletitle{Evaluating large language models trained on code}.
\newblock \bibinfo{journal}{\emph{arXiv preprint arXiv:2107.03374}} (\bibinfo{year}{2021}).
\newblock


\bibitem[Cipriano and Alves(2023)]%
        {cipriano2023gpt}
\bibfield{author}{\bibinfo{person}{Bruno~Pereira Cipriano} {and} \bibinfo{person}{Pedro Alves}.} \bibinfo{year}{2023}\natexlab{}.
\newblock \showarticletitle{GPT-3 vs Object Oriented Programming Assignments: An Experience Report}. In \bibinfo{booktitle}{\emph{Proceedings of the 2023 Conference on Innovation and Technology in Computer Science Education V. 1}}. \bibinfo{pages}{61--67}.
\newblock


\bibitem[Crow et~al\mbox{.}(2018)]%
        {crow2018intelligent}
\bibfield{author}{\bibinfo{person}{Tyne Crow}, \bibinfo{person}{Andrew Luxton-Reilly}, {and} \bibinfo{person}{Burkhard Wuensche}.} \bibinfo{year}{2018}\natexlab{}.
\newblock \showarticletitle{Intelligent tutoring systems for programming education: a systematic review}. In \bibinfo{booktitle}{\emph{Proceedings of the 20th Australasian Computing Education Conference}}. \bibinfo{pages}{53--62}.
\newblock


\bibitem[Demana and Waits(2000)]%
        {demana2000calculators}
\bibfield{author}{\bibinfo{person}{Franklin Demana} {and} \bibinfo{person}{BK Waits}.} \bibinfo{year}{2000}\natexlab{}.
\newblock \showarticletitle{Calculators in mathematics teaching and learning}.
\newblock \bibinfo{journal}{\emph{Past, present, and future. In Learning Mathematics for a New Century}} (\bibinfo{year}{2000}), \bibinfo{pages}{51--66}.
\newblock


\bibitem[Denny et~al\mbox{.}(2023a)]%
        {denny2023chat}
\bibfield{author}{\bibinfo{person}{Paul Denny}, \bibinfo{person}{Brett~A Becker}, \bibinfo{person}{Juho Leinonen}, {and} \bibinfo{person}{James Prather}.} \bibinfo{year}{2023}\natexlab{a}.
\newblock \showarticletitle{Chat Overflow: Artificially Intelligent Models for Computing Education-renAIssance or apocAIypse?}. In \bibinfo{booktitle}{\emph{Proceedings of the 2023 Conference on Innovation and Technology in Computer Science Education V. 1}}. \bibinfo{pages}{3--4}.
\newblock


\bibitem[Denny et~al\mbox{.}(2024)]%
        {denny2024desirable}
\bibfield{author}{\bibinfo{person}{Paul Denny}, \bibinfo{person}{Stephen MacNeil}, \bibinfo{person}{Jaromir Savelka}, \bibinfo{person}{Leo Porter}, {and} \bibinfo{person}{Andrew Luxton-Reilly}.} \bibinfo{year}{2024}\natexlab{}.
\newblock \showarticletitle{Desirable Characteristics for AI Teaching Assistants in Programming Education}. In \bibinfo{booktitle}{\emph{Proceedings of the 2024 on Innovation and Technology in Computer Science Education V. 1}} (Milan, Italy) \emph{(\bibinfo{series}{ITiCSE 2024})}. \bibinfo{publisher}{Association for Computing Machinery}, \bibinfo{address}{New York, NY, USA}, \bibinfo{pages}{408–414}.
\newblock
\showISBNx{9798400706004}
\urldef\tempurl%
\url{https://doi.org/10.1145/3649217.3653574}
\showDOI{\tempurl}


\bibitem[Denny et~al\mbox{.}(2023b)]%
        {denny2023computing}
\bibfield{author}{\bibinfo{person}{Paul Denny}, \bibinfo{person}{James Prather}, \bibinfo{person}{Brett~A Becker}, \bibinfo{person}{James Finnie-Ansley}, \bibinfo{person}{Arto Hellas}, \bibinfo{person}{Juho Leinonen}, \bibinfo{person}{Andrew Luxton-Reilly}, \bibinfo{person}{Brent~N Reeves}, \bibinfo{person}{Eddie~Antonio Santos}, {and} \bibinfo{person}{Sami Sarsa}.} \bibinfo{year}{2023}\natexlab{b}.
\newblock \showarticletitle{Computing Education in the Era of Generative AI}.
\newblock \bibinfo{journal}{\emph{arXiv preprint arXiv:2306.02608}} (\bibinfo{year}{2023}).
\newblock


\bibitem[Finnie-Ansley et~al\mbox{.}(2022)]%
        {finnie2022robots}
\bibfield{author}{\bibinfo{person}{James Finnie-Ansley}, \bibinfo{person}{Paul Denny}, \bibinfo{person}{Brett~A Becker}, \bibinfo{person}{Andrew Luxton-Reilly}, {and} \bibinfo{person}{James Prather}.} \bibinfo{year}{2022}\natexlab{}.
\newblock \showarticletitle{The robots are coming: Exploring the implications of openai codex on introductory programming}. In \bibinfo{booktitle}{\emph{Proceedings of the 24th Australasian Computing Education Conference}}. \bibinfo{pages}{10--19}.
\newblock


\bibitem[Finnie-Ansley et~al\mbox{.}(2023)]%
        {finnie2023my}
\bibfield{author}{\bibinfo{person}{James Finnie-Ansley}, \bibinfo{person}{Paul Denny}, \bibinfo{person}{Andrew Luxton-Reilly}, \bibinfo{person}{Eddie~Antonio Santos}, \bibinfo{person}{James Prather}, {and} \bibinfo{person}{Brett~A Becker}.} \bibinfo{year}{2023}\natexlab{}.
\newblock \showarticletitle{My AI Wants to Know if This Will Be on the Exam: Testing OpenAI’s Codex on CS2 Programming Exercises}. In \bibinfo{booktitle}{\emph{Proceedings of the 25th Australasian Computing Education Conference}}. \bibinfo{pages}{97--104}.
\newblock


\bibitem[Hellas et~al\mbox{.}(2023)]%
        {hellas2023exploring}
\bibfield{author}{\bibinfo{person}{Arto Hellas}, \bibinfo{person}{Juho Leinonen}, \bibinfo{person}{Sami Sarsa}, \bibinfo{person}{Charles Koutcheme}, \bibinfo{person}{Lilja Kujanp{\"a}{\"a}}, {and} \bibinfo{person}{Juha Sorva}.} \bibinfo{year}{2023}\natexlab{}.
\newblock \showarticletitle{Exploring the Responses of Large Language Models to Beginner Programmers' Help Requests}. In \bibinfo{booktitle}{\emph{Proceedings of the 2023 ACM Conference on International Computing Education Research V.1}}.
\newblock


\bibitem[Kazemitabaar et~al\mbox{.}(2023)]%
        {kazemitabaar2023studying}
\bibfield{author}{\bibinfo{person}{Majeed Kazemitabaar}, \bibinfo{person}{Justin Chow}, \bibinfo{person}{Carl Ka~To Ma}, \bibinfo{person}{Barbara~J Ericson}, \bibinfo{person}{David Weintrop}, {and} \bibinfo{person}{Tovi Grossman}.} \bibinfo{year}{2023}\natexlab{}.
\newblock \showarticletitle{Studying the effect of AI Code Generators on Supporting Novice Learners in Introductory Programming}. In \bibinfo{booktitle}{\emph{Proceedings of the 2023 CHI Conference on Human Factors in Computing Systems}}. \bibinfo{pages}{1--23}.
\newblock


\bibitem[Kazemitabaar et~al\mbox{.}(2024)]%
        {kazemi2024codeaid}
\bibfield{author}{\bibinfo{person}{Majeed Kazemitabaar}, \bibinfo{person}{Runlong Ye}, \bibinfo{person}{Xiaoning Wang}, \bibinfo{person}{Austin~Zachary Henley}, \bibinfo{person}{Paul Denny}, \bibinfo{person}{Michelle Craig}, {and} \bibinfo{person}{Tovi Grossman}.} \bibinfo{year}{2024}\natexlab{}.
\newblock \showarticletitle{CodeAid: Evaluating a Classroom Deployment of an LLM-based Programming Assistant that Balances Student and Educator Needs}. In \bibinfo{booktitle}{\emph{Proceedings of the CHI Conference on Human Factors in Computing Systems}} \emph{(\bibinfo{series}{CHI '24})}.
\newblock


\bibitem[Leinonen et~al\mbox{.}(2023a)]%
        {leinonen2023comparing}
\bibfield{author}{\bibinfo{person}{Juho Leinonen}, \bibinfo{person}{Paul Denny}, \bibinfo{person}{Stephen MacNeil}, \bibinfo{person}{Sami Sarsa}, \bibinfo{person}{Seth Bernstein}, \bibinfo{person}{Joanne Kim}, \bibinfo{person}{Andrew Tran}, {and} \bibinfo{person}{Arto Hellas}.} \bibinfo{year}{2023}\natexlab{a}.
\newblock \showarticletitle{Comparing Code Explanations Created by Students and Large Language Models}. In \bibinfo{booktitle}{\emph{Proceedings of the 2023 Conference on Innovation and Technology in Computer Science Education V. 1}}. \bibinfo{publisher}{ACM}, \bibinfo{pages}{124–130}.
\newblock
\showISBNx{9798400701382}


\bibitem[Leinonen et~al\mbox{.}(2023b)]%
        {leinonen2023using}
\bibfield{author}{\bibinfo{person}{Juho Leinonen}, \bibinfo{person}{Arto Hellas}, \bibinfo{person}{Sami Sarsa}, \bibinfo{person}{Brent Reeves}, \bibinfo{person}{Paul Denny}, \bibinfo{person}{James Prather}, {and} \bibinfo{person}{Brett~A Becker}.} \bibinfo{year}{2023}\natexlab{b}.
\newblock \showarticletitle{Using Large Language Models to Enhance Programming Error Messages}. In \bibinfo{booktitle}{\emph{Proceedings of the 54th ACM Technical Symposium on Computer Science Education V. 1.}} ACM.
\newblock


\bibitem[Liffiton et~al\mbox{.}(2023)]%
        {liffiton2023codehelp}
\bibfield{author}{\bibinfo{person}{Mark Liffiton}, \bibinfo{person}{Brad Sheese}, \bibinfo{person}{Jaromir Savelka}, {and} \bibinfo{person}{Paul Denny}.} \bibinfo{year}{2023}\natexlab{}.
\newblock \bibinfo{title}{CodeHelp: Using Large Language Models with Guardrails for Scalable Support in Programming Classes}.
\newblock
\newblock
\showeprint[arxiv]{2308.06921}~[cs.CY]


\bibitem[Liu et~al\mbox{.}(2024a)]%
        {harvardcs50}
\bibfield{author}{\bibinfo{person}{Rongxin Liu}, \bibinfo{person}{Carter Zenke}, \bibinfo{person}{Charlie Liu}, \bibinfo{person}{Andrew Holmes}, \bibinfo{person}{Patrick Thornton}, {and} \bibinfo{person}{David~J. Malan}.} \bibinfo{year}{2024}\natexlab{a}.
\newblock \showarticletitle{Teaching CS50 with AI: Leveraging Generative Artificial Intelligence in Computer Science Education}. In \bibinfo{booktitle}{\emph{Proceedings of the 55th ACM Technical Symposium on Computer Science Education V. 1}} (Portland, OR, USA) \emph{(\bibinfo{series}{SIGCSE 2024})}. \bibinfo{publisher}{Association for Computing Machinery}, \bibinfo{address}{New York, NY, USA}, \bibinfo{pages}{750–756}.
\newblock
\showISBNx{9798400704239}
\urldef\tempurl%
\url{https://doi.org/10.1145/3626252.3630938}
\showDOI{\tempurl}


\bibitem[Liu et~al\mbox{.}(2024b)]%
        {liu2024teaching}
\bibfield{author}{\bibinfo{person}{Rongxin Liu}, \bibinfo{person}{Carter Zenke}, \bibinfo{person}{Charlie Liu}, \bibinfo{person}{Andrew Holmes}, \bibinfo{person}{Patrick Thornton}, {and} \bibinfo{person}{David~J. Malan}.} \bibinfo{year}{2024}\natexlab{b}.
\newblock \showarticletitle{Teaching CS50 with AI: Leveraging Generative Artificial Intelligence in Computer Science Education}. In \bibinfo{booktitle}{\emph{Proceedings of the 55th ACM Technical Symposium on Computer Science Education V. 2}} (Portland, OR, USA) \emph{(\bibinfo{series}{SIGCSE 2024})}. \bibinfo{publisher}{Association for Computing Machinery}, \bibinfo{address}{New York, NY, USA}, \bibinfo{pages}{1927}.
\newblock
\showISBNx{9798400704246}


\bibitem[MacNeil et~al\mbox{.}(2023a)]%
        {macneil2023implications}
\bibfield{author}{\bibinfo{person}{Stephen MacNeil}, \bibinfo{person}{Joanne Kim}, \bibinfo{person}{Juho Leinonen}, \bibinfo{person}{Paul Denny}, \bibinfo{person}{Seth Bernstein}, \bibinfo{person}{Brett~A Becker}, \bibinfo{person}{Michel Wermelinger}, \bibinfo{person}{Arto Hellas}, \bibinfo{person}{Andrew Tran}, \bibinfo{person}{Sami Sarsa}, {et~al\mbox{.}}} \bibinfo{year}{2023}\natexlab{a}.
\newblock \showarticletitle{The Implications of Large Language Models for CS Teachers and Students}. In \bibinfo{booktitle}{\emph{Proc. of the 54th ACM Technical Symposium on Computer Science Education}}, Vol.~\bibinfo{volume}{2}.
\newblock


\bibitem[MacNeil et~al\mbox{.}(2023b)]%
        {macneil2023experiences}
\bibfield{author}{\bibinfo{person}{Stephen MacNeil}, \bibinfo{person}{Andrew Tran}, \bibinfo{person}{Arto Hellas}, \bibinfo{person}{Joanne Kim}, \bibinfo{person}{Sami Sarsa}, \bibinfo{person}{Paul Denny}, \bibinfo{person}{Seth Bernstein}, {and} \bibinfo{person}{Juho Leinonen}.} \bibinfo{year}{2023}\natexlab{b}.
\newblock \showarticletitle{Experiences from using code explanations generated by large language models in a web software development e-book}. In \bibinfo{booktitle}{\emph{Proceedings of the 54th ACM Technical Symposium on Computer Science Education V. 1}}. \bibinfo{pages}{931--937}.
\newblock


\bibitem[Markel and Guo(2021)]%
        {markel2021inside}
\bibfield{author}{\bibinfo{person}{Julia~M Markel} {and} \bibinfo{person}{Philip~J Guo}.} \bibinfo{year}{2021}\natexlab{}.
\newblock \showarticletitle{Inside the mind of a CS undergraduate TA: A firsthand account of undergraduate peer tutoring in computer labs}. In \bibinfo{booktitle}{\emph{Proceedings of the 52nd ACM Technical Symposium on Computer Science Education}}. \bibinfo{pages}{502--508}.
\newblock


\bibitem[Mirhosseini et~al\mbox{.}(2023)]%
        {mirhosseini2023your}
\bibfield{author}{\bibinfo{person}{Samim Mirhosseini}, \bibinfo{person}{Austin~Z Henley}, {and} \bibinfo{person}{Chris Parnin}.} \bibinfo{year}{2023}\natexlab{}.
\newblock \showarticletitle{What is your biggest pain point? an investigation of cs instructor obstacles, workarounds, and desires}. In \bibinfo{booktitle}{\emph{Proceedings of the 54th ACM Technical Symposium on Computer Science Education V. 1}}. \bibinfo{pages}{291--297}.
\newblock


\bibitem[Mitra et~al\mbox{.}(2024)]%
        {mitra2024elevating}
\bibfield{author}{\bibinfo{person}{Chancharik Mitra}, \bibinfo{person}{Mihran Miroyan}, \bibinfo{person}{Rishi Jain}, \bibinfo{person}{Vedant Kumud}, \bibinfo{person}{Gireeja Ranade}, {and} \bibinfo{person}{Narges Norouzi}.} \bibinfo{year}{2024}\natexlab{}.
\newblock \showarticletitle{Elevating Learning Experiences: Leveraging Large Language Models as Student-Facing Assistants in Discussion Forums}. In \bibinfo{booktitle}{\emph{Proceedings of the 55th ACM Technical Symposium on Computer Science Education V. 2}} (Portland, OR, USA) \emph{(\bibinfo{series}{SIGCSE 2024})}.
\newblock


\bibitem[Ofgang(2023)]%
        {ofgang2023khanmigo}
\bibfield{author}{\bibinfo{person}{Erik Ofgang}.} \bibinfo{year}{2023}\natexlab{}.
\newblock \showarticletitle{What is Khanmigo? The GPT-4 learning tool explained by Sal Khan}.
\newblock \bibinfo{journal}{\emph{Tech \& Learn}} (\bibinfo{year}{2023}).
\newblock


\bibitem[Pardos and Bhandari(2023)]%
        {pardos2023learning}
\bibfield{author}{\bibinfo{person}{Zachary~A Pardos} {and} \bibinfo{person}{Shreya Bhandari}.} \bibinfo{year}{2023}\natexlab{}.
\newblock \showarticletitle{Learning gain differences between ChatGPT and human tutor generated algebra hints}.
\newblock \bibinfo{journal}{\emph{arXiv preprint arXiv:2302.06871}} (\bibinfo{year}{2023}).
\newblock


\bibitem[Piech et~al\mbox{.}(2012)]%
        {piech2012modeling}
\bibfield{author}{\bibinfo{person}{Chris Piech}, \bibinfo{person}{Mehran Sahami}, \bibinfo{person}{Daphne Koller}, \bibinfo{person}{Steve Cooper}, {and} \bibinfo{person}{Paulo Blikstein}.} \bibinfo{year}{2012}\natexlab{}.
\newblock \showarticletitle{Modeling how students learn to program}. In \bibinfo{booktitle}{\emph{Proceedings of the 43rd ACM technical symposium on Computer Science Education}}. \bibinfo{pages}{153--160}.
\newblock


\bibitem[Prather et~al\mbox{.}(2024)]%
        {prather2024instructors}
\bibfield{author}{\bibinfo{person}{James Prather}, \bibinfo{person}{Juho Leinonen}, \bibinfo{person}{Natalie Kiesler}, \bibinfo{person}{Jamie~Gorson Benario}, \bibinfo{person}{Sam Lau}, \bibinfo{person}{Stephen MacNeil}, \bibinfo{person}{Narges Norouzi}, \bibinfo{person}{Simone Opel}, \bibinfo{person}{Virginia Pettit}, \bibinfo{person}{Leo Porter}, {et~al\mbox{.}}} \bibinfo{year}{2024}\natexlab{}.
\newblock \showarticletitle{How Instructors Incorporate Generative AI into Teaching Computing}.
\newblock In \bibinfo{booktitle}{\emph{Proceedings of the 2024 on Innovation and Technology in Computer Science Education V. 2}}. \bibinfo{pages}{771--772}.
\newblock


\bibitem[Prather et~al\mbox{.}(2023)]%
        {prather2023s}
\bibfield{author}{\bibinfo{person}{James Prather}, \bibinfo{person}{Brent~N Reeves}, \bibinfo{person}{Paul Denny}, \bibinfo{person}{Brett~A Becker}, \bibinfo{person}{Juho Leinonen}, \bibinfo{person}{Andrew Luxton-Reilly}, \bibinfo{person}{Garrett Powell}, \bibinfo{person}{James Finnie-Ansley}, {and} \bibinfo{person}{Eddie~Antonio Santos}.} \bibinfo{year}{2023}\natexlab{}.
\newblock \showarticletitle{" It's Weird That it Knows What I Want": Usability and Interactions with Copilot for Novice Programmers}.
\newblock \bibinfo{journal}{\emph{arXiv preprint arXiv:2304.02491}} (\bibinfo{year}{2023}).
\newblock


\bibitem[Reeves et~al\mbox{.}(2023)]%
        {reeves2023evaluating}
\bibfield{author}{\bibinfo{person}{Brent Reeves}, \bibinfo{person}{Sami Sarsa}, \bibinfo{person}{James Prather}, \bibinfo{person}{Paul Denny}, \bibinfo{person}{Brett~A Becker}, \bibinfo{person}{Arto Hellas}, \bibinfo{person}{Bailey Kimmel}, \bibinfo{person}{Garrett Powell}, {and} \bibinfo{person}{Juho Leinonen}.} \bibinfo{year}{2023}\natexlab{}.
\newblock \showarticletitle{Evaluating the Performance of Code Generation Models for Solving Parsons Problems With Small Prompt Variations}. In \bibinfo{booktitle}{\emph{Proceedings of the 2023 Conference on Innovation and Technology in Computer Science Education V. 1}}. \bibinfo{pages}{299--305}.
\newblock


\bibitem[Sarsa et~al\mbox{.}(2022)]%
        {sarsa2022automatic}
\bibfield{author}{\bibinfo{person}{Sami Sarsa}, \bibinfo{person}{Paul Denny}, \bibinfo{person}{Arto Hellas}, {and} \bibinfo{person}{Juho Leinonen}.} \bibinfo{year}{2022}\natexlab{}.
\newblock \showarticletitle{Automatic generation of programming exercises and code explanations using large language models}. In \bibinfo{booktitle}{\emph{Proceedings of the 2022 ACM Conference on International Computing Education Research-Volume 1}}. \bibinfo{pages}{27--43}.
\newblock


\bibitem[Sheese et~al\mbox{.}(2024)]%
        {sheese2024patterns}
\bibfield{author}{\bibinfo{person}{Brad Sheese}, \bibinfo{person}{Mark Liffiton}, \bibinfo{person}{Jaromir Savelka}, {and} \bibinfo{person}{Paul Denny}.} \bibinfo{year}{2024}\natexlab{}.
\newblock \showarticletitle{Patterns of Student Help-Seeking When Using a Large Language Model-Powered Programming Assistant}. In \bibinfo{booktitle}{\emph{Proceedings of the 26th Australasian Computing Education Conference}}. \bibinfo{pages}{49--57}.
\newblock


\bibitem[Soloway et~al\mbox{.}(1994)]%
        {soloway1994learner}
\bibfield{author}{\bibinfo{person}{Elliot Soloway}, \bibinfo{person}{Mark Guzdial}, {and} \bibinfo{person}{Kenneth~E. Hay}.} \bibinfo{year}{1994}\natexlab{}.
\newblock \showarticletitle{Learner-centered design: the challenge for HCI in the 21st century}.
\newblock \bibinfo{journal}{\emph{Interactions}}  \bibinfo{volume}{1} (\bibinfo{year}{1994}), \bibinfo{pages}{36--48}.
\newblock


\bibitem[Suzuki et~al\mbox{.}(2017)]%
        {suzuki2017tracediff}
\bibfield{author}{\bibinfo{person}{Ryo Suzuki}, \bibinfo{person}{Gustavo Soares}, \bibinfo{person}{Andrew Head}, \bibinfo{person}{Elena Glassman}, \bibinfo{person}{Ruan Reis}, \bibinfo{person}{Melina Mongiovi}, \bibinfo{person}{Loris D'Antoni}, {and} \bibinfo{person}{Bjoern Hartmann}.} \bibinfo{year}{2017}\natexlab{}.
\newblock \showarticletitle{Tracediff: Debugging unexpected code behavior using trace divergences}. In \bibinfo{booktitle}{\emph{2017 IEEE Symposium on Visual Languages and Human-Centric Computing (VL/HCC)}}. IEEE, \bibinfo{pages}{107--115}.
\newblock


\bibitem[Wang et~al\mbox{.}(2024)]%
        {wang2024large}
\bibfield{author}{\bibinfo{person}{Sierra Wang}, \bibinfo{person}{John Mitchell}, {and} \bibinfo{person}{Chris Piech}.} \bibinfo{year}{2024}\natexlab{}.
\newblock \showarticletitle{A Large Scale RCT on Effective Error Messages in CS1}. In \bibinfo{booktitle}{\emph{Proceedings of the 55th ACM Technical Symposium on Computer Science Education V. 1}} (Portland, OR, USA) \emph{(\bibinfo{series}{SIGCSE 2024})}. \bibinfo{pages}{1395–1401}.
\newblock


\bibitem[Wu et~al\mbox{.}(2022)]%
        {wu2022student}
\bibfield{author}{\bibinfo{person}{Yu-Chieh Wu}, \bibinfo{person}{Andrew Petersen}, {and} \bibinfo{person}{Lisa Zhang}.} \bibinfo{year}{2022}\natexlab{}.
\newblock \showarticletitle{Student Reactions to Bots on Course Q\&A Platform}. In \bibinfo{booktitle}{\emph{Proceedings of the 27th ACM Conference on on Innovation and Technology in Computer Science Education Vol. 2}}. \bibinfo{pages}{621--621}.
\newblock


\bibitem[Yuan et~al\mbox{.}(2023)]%
        {yuan2023evaluating}
\bibfield{author}{\bibinfo{person}{Zhiqiang Yuan}, \bibinfo{person}{Junwei Liu}, \bibinfo{person}{Qiancheng Zi}, \bibinfo{person}{Mingwei Liu}, \bibinfo{person}{Xin Peng}, {and} \bibinfo{person}{Yiling Lou}.} \bibinfo{year}{2023}\natexlab{}.
\newblock \showarticletitle{Evaluating instruction-tuned large language models on code comprehension and generation}.
\newblock \bibinfo{journal}{\emph{arXiv preprint arXiv:2308.01240}} (\bibinfo{year}{2023}).
\newblock


\end{thebibliography}

\end{document}